\documentclass[letterpaper]{article} % DO NOT CHANGE THIS
\usepackage{aaai25}  % DO NOT CHANGE THIS
\usepackage{times}  % DO NOT CHANGE THIS
\usepackage{helvet}  % DO NOT CHANGE THIS
\usepackage{courier}  % DO NOT CHANGE THIS
\usepackage[hyphens]{url}  % DO NOT CHANGE THIS
\usepackage{graphicx} % DO NOT CHANGE THIS
\usepackage{natbib}  % DO NOT CHANGE THIS AND DO NOT ADD ANY OPTIONS TO IT
\usepackage{caption} % DO NOT CHANGE THIS AND DO NOT ADD ANY OPTIONS TO IT
\usepackage{algorithm}
\usepackage{algorithmic}
\usepackage{xcolor}
\usepackage{makecell}
\usepackage{multicol}
\usepackage{booktabs}
\usepackage{multirow}
\usepackage{xspace}
\usepackage{amsmath}

\usepackage{etoolbox}

\newcommand{\IPAs}[0]{\textsc{IPAs}\xspace}
\newcommand{\IPA}[0]{\textsc{IPA}\xspace}
\newcommand{\benchmark}[0]{\textsc{C-Pack-IPAs}\xspace}
\newcommand{\Verifix}[0]{\textsc{Ve\-ri\-fix}\xspace}
\newcommand{\Clara}[0]{\textsc{Cla\-ra}\xspace}
\newcommand{\LLMs}[0]{\textsc{LLMs}\xspace}
\newcommand{\LLM}[0]{\textsc{LLM}\xspace}
\newcommand{\LLMCs}[0]{\textsc{LLMCs}\xspace}
\newcommand{\LLMC}[0]{\textsc{LLMC}\xspace}

\newcommand\AST[0]{\textsc{AST}\xspace}
\newcommand\ASTs[0]{\textsc{ASTs}\xspace}
\newcommand{\CEGIS}[0]{\textsc{CEGIS}\xspace}

\newcommand{\Granite}[0]{\textsc{Gra\-ni\-te}\xspace}
\newcommand{\Gemma}[0]{\textsc{Gem\-ma}\xspace}
\newcommand{\CodeGemma}[0]{\textsc{Co\-de\-Gem\-ma}\xspace}
\newcommand{\Llama}[0]{\textsc{Lla\-ma\-3}\xspace}
\newcommand{\CodeLlama}[0]{\textsc{Co\-de\-Lla\-ma}\xspace}
\newcommand{\lPhi}[0]{\textsc{Phi\-3}\xspace}

\DeclareMathOperator{\ted}{TED}

\usepackage[newfloat,frozencache,cachedir=.]{minted}
\newenvironment{code}{\floatstyle{plaintop}%
\captionsetup{type=listing, labelfont=bf,justification=raggedright, singlelinecheck=false,skip=0pt}}{}
\SetupFloatingEnvironment{listing}{name=\textbf{Listing}}

\usepackage{tikz}
\usetikzlibrary{arrows,automata,backgrounds,calc,chains,external,fit,graphs,patterns,positioning,shapes,shapes.geometric}

\tikzset{
    >=stealth',
    punkt/.style={
           rectangle,
           rounded corners,
           draw=black, very thick,
           text width=6.5em,
           minimum height=2em,
           text centered},
    pil/.style={
           ->,
           thick,
           shorten <=2pt,
           shorten >=2pt,}
}

\tikzset{fancy/.style={rectangle,
		rounded corners=1mm,
		ultra thin,
		draw=white,
		top color=white,
		bottom color=black!20,
		draw}}

\definecolor{highClr}{rgb}{1.0, 1.0, 0.0}

\colorlet{edgeClr}{orange!80!black}
\tikzset{sandEdge/.style={
		>=stealth,
		shorten >=1pt,
		thick,
		bend left,
		text=black,
		edgeClr,
	}}

\tikzset{fadedEdge/.style={
		->,
		>=stealth,
		shorten >=1pt,
		thick,
		edgeClr!20,
	}}

\tikzset{weigthLabel/.style={
		text=black,
		sloped,
		midway,
		}}

\tikzset{fadedWeigth/.style={
		text=lightgray!40,
		sloped,
		midway,
		anchor=south,
		}}

\tikzset{blueVertex/.style={
		rectangle,minimum size=6mm,rounded corners=3mm,
		top color=white,bottom color=blue!35!cyan!25!,
		font=\ttfamily,
		text=black,
	}}

\tikzset{blueVertexG/.style={
		rectangle,minimum size=6mm,rounded corners=3mm,
		top color=white,bottom color=blue!35!cyan!25!,
		font=\ttfamily,
		text=black,
		draw=green,
		thick,
	},
	blueVertexY/.style={
		rectangle,minimum size=6mm,rounded corners=3mm,
		top color=white,bottom color=blue!35!cyan!25!,
		font=\ttfamily,
		text=black,
		draw=yellow,
		thick,
	},
	blueVertexO/.style={
		rectangle,minimum size=6mm,rounded corners=3mm,
		top color=white,bottom color=blue!35!cyan!25!,
		font=\ttfamily,
		text=black,
		draw=orange,
		thick,
	}}

\colorlet{noteClr}{lightgray!30!white!50}

\tikzset{noteBckg/.style={
		rounded corners=8pt,fill=noteClr,
	},
	noteStl/.style={
		font=\scriptsize,
		align=center,
		text=black
	}}

\makeatletter

\newcommand*\mysize{%
  \@setfontsize\mysize{8}{8}%
}
\makeatother
\setminted{fontsize=\mysize}

\title{Counterexample Guided Program Repair Using Zero-Shot Learning and MaxSAT-based Fault Localization}

\author{
    Pedro Orvalho\textsuperscript{\rm 1}\thanks{Part of this work was conducted at INESC-ID, IST, UL.},
    Mikoláš Janota\textsuperscript{\rm 2},
    Vasco Manquinho\textsuperscript{\rm 3}
}
\affiliations{
    \textsuperscript{\rm 1}Department of Computer Science, University of Oxford, Oxford, UK\\
    \textsuperscript{\rm 2}CIIRC, Czech Technical University in Prague, Czechia\\
    \textsuperscript{\rm 3}INESC-ID, IST, Universidade de Lisboa, Portugal\\
    pmorvalho@gmail.com, mikolas.janota@cvut.cz, vasco.manquinho@tecnico.ulisboa.pt

}

\begin{document}

\maketitle

\begin{abstract}

Automated Program Repair (APR) for introductory programming assignments (\IPAs) is motivated by the large number of student enrollments in programming courses each year. Since providing feedback on programming assignments requires substantial time and effort from faculty, personalized automated feedback often involves suggesting repairs to students' programs. Symbolic semantic repair approaches, which rely on Formal Methods (FM), check a program's  execution against a test suite or reference solution, are effective but limited. These tools excel at identifying buggy parts but can only fix programs if the correct implementation and the faulty one share the same control flow graph. Conversely, Large Language Models (\LLMs) are used for program repair but often make extensive rewrites instead of minimal adjustments. This tends to lead to more invasive fixes, making it harder for students to learn from their mistakes.
In summary, \LLMs excel at completing strings, while FM-based fault localization excel at identifying buggy parts of a program.

In this paper, we propose a novel approach that combines the strengths of both FM-based fault localization and \LLMs, via zero-shot learning, to enhance APR for \IPAs. Our method uses MaxSAT-based fault localization to identify buggy parts of a program, then presents the \LLM with a program sketch devoid of these buggy statements. This hybrid approach follows a Counterexample Guided Inductive Synthesis (\CEGIS) loop to iteratively refine the program. We ask the \LLM to synthesize the missing parts, which are then checked against a test suite. If the suggested program is incorrect, a counterexample from the test suite is fed back to the \LLM for revised synthesis.
Our experiments on 1,431 incorrect student programs show that our counterexample guided approach, using MaxSAT-based bug-free program sketches, significantly improves the repair capabilities of all six evaluated \LLMs. This method allows \LLMs to repair more programs and produce smaller fixes, outperforming other configurations and state-of-the-art symbolic program repair tools.
\end{abstract}


\section{Introduction}

Every year, thousands of students enroll in programming-oriented courses. With the rapid growth of Computer Science courses, providing personalized and timely feedback on \emph{introductory programming assignments} (\IPAs) and software projects has become a significant challenge, requiring substantial time and effort from faculty~\cite{GitSEED-sigcse-virtual24, fse22-multIPAs}.

\emph{Automated Program Repair} (APR) has emerged as a promising solution to this challenge, aiming to deliver automated, comprehensive, and personalized feedback to students about their programming errors~\cite{clara, verifix, sarfgen, refactory}.
Traditional semantic APR techniques based on \emph{Formal Methods} (FM), while providing high-quality fixes, are often slow and may struggle when the correct implementation diverges significantly from the erroneous one~\cite{its22-improving-clara-matching-procedure}.
% \todo{cite more}
These APR approaches do not guarantee minimal repairs, as they align an incorrect submission with a correct implementation for the same \IPA. If the alignment is not possible, these tools return a structural mismatch error, leaving the program unrepaired.
In the past decade, there has been a surge in Machine Learning (ML) techniques for APR~\cite{deepfix, deepDelta, rlassist, drRepair, refazer, sk_p,synFix,ecai23-GNNs-4-var-mapping}.
ML-based approaches require multiple correct implementations to generate high-quality repairs, and need considerable time and resources to train on correct programs.
While these approaches generate repairs more quickly, they often produce imprecise and non-minimal fixes~\cite{sarfgen}.

More recently, \emph{Large Language Models} (\LLMs) trained on code (\LLMCs) have shown great potential in generating program fixes~\cite{aaai23-repair-multilingual-LLMs,ase23-LLMs-plastic-surgery,fse23-inferix,fse23-copiloting-copilots,icse23-APR-LLMs,icse23-APR-pretrained-LLMs,oopsla24-PyDex,edm23-feedback-syntax-errors-LLMs}.
\LLM-based APR can be performed using zero-shot learning~\cite{fse22-APR-zero-shot-learning}, few-shot learning~\cite{oopsla24-PyDex} or fine-tuned models~\cite{fse23-inferix}. Fine-tuned models are the most commonly used, where the model is trained for a specific task. Conversely, zero-shot learning refers to the ability of a model to correctly perform a task without having seen any examples of that task during training. Few-shot learning refers to the \LLMs's ability to perform tasks correctly with only a small number of examples provided.
Furthermore, the ability to generalize using zero or few-shot learning enables \LLMs to handle a wide range of tasks without the need for costly retraining or fine-tuning. Nonetheless, few-shot learning can lead to larger fixes than necessary, as it is based on a limited number of examples.
\LLMs do not guarantee minimal repairs and typically rewrite most of the student's implementation to fix it, rather than making minimal adjustments, making their fixes less efficient and harder for students to learn from.

\begin{table*}[t!]
\hfill
\begin{minipage}[t!]{0.5\linewidth}
\centering
\begin{code}
\caption{Semantically incorrect program. Faulty lines: \{4,8\}.}%
\label{code:motivating_eg}
\begin{minted}[escapeinside=||,tabsize=1,obeytabs,xleftmargin=2pt,linenos]{C}
int main(){ // finds maximum of 3 numbers
   int f,s,t;
   scanf("%d%d%d",&f,&s,&t);
   if (f < s && f >= t) //fix: f >= s
      printf("%d",f);
   else if (s > f && s >= t)
      printf("%d",s);
   else if (t < f && t < s) //fix: t > f and t > s
      printf("%d",t);

   return 0;
}
\end{minted}
\end{code}
% \vspace{-0.3in}
\end{minipage}
\hfill
\begin{minipage}[t!]{0.45\linewidth}
\centering
\begin{code}
\vspace{-.8cm}
\caption{Reference implementation.}%
\label{code:motivating_ref}
\begin{minted}[escapeinside=||,tabsize=1,obeytabs,xleftmargin=2pt,linenos]{C}
int main() {
   int m1,m2,m3,m;
   scanf("%d%d%d",&m1,&m2,&m3);
   m = m1 > m2 ? m1 : m2;
   m = m3 > m ? m3 : m;
   printf("%d\n", m);

   return 0;
}
\end{minted}
\end{code}
\end{minipage}

\hfill
\begin{minipage}[t!]{0.3\linewidth}
% \vspace{-0.2in}
\centering
\vspace{0.1in}
\begin{code}
%\caption{Program sketch with holes instead of the initial buggy statements.}%
\caption{Program sketch with holes.}%
\label{code:motivating-sketch}
\begin{minted}[escapeinside=||,tabsize=1,obeytabs,xleftmargin=2pt,linenos]{C}
int main(){
   int f,s,t;
   scanf("%d%d%d",&f,&s,&t);
   @ HOLE 1 @
      printf("%d",f);
   else if (s > f && s >= t)
      printf("%d",s);
   @ HOLE 2 @
      printf("%d",t);

   return 0;
}
\end{minted}
\end{code}
\end{minipage}
% \hfill
\begin{minipage}[t!]{0.19\linewidth}
\vspace{0.2in}
\centering
\resizebox{0.75\linewidth}{!}{%
\begin{tabular}{clll}
\cline{2-4}
\multicolumn{1}{l|}{}  & \multicolumn{3}{c|}{\textbf{Tests}}                                      \\ \cline{2-4}
\multicolumn{1}{l|}{}                                 & \multicolumn{1}{l|}{\textit{t0}} & \multicolumn{1}{l|}{\textit{t1}} & \multicolumn{1}{l|}{\textit{t2}} \\ \hline
\multicolumn{1}{|c|}{\multirow{3}{*}{\textbf{Input}}} & \multicolumn{1}{l|}{1}           & \multicolumn{1}{l|}{6}           & \multicolumn{1}{l|}{-1}          \\ \cline{2-4}
\multicolumn{1}{|c|}{} & \multicolumn{1}{l|}{2} & \multicolumn{1}{l|}{2} & \multicolumn{1}{l|}{3} \\ \cline{2-4}
\multicolumn{1}{|c|}{} & \multicolumn{1}{l|}{3} & \multicolumn{1}{l|}{1} & \multicolumn{1}{l|}{1} \\ \hline
\multicolumn{1}{l}{}   &                        &                        &                        \\ \hline
\multicolumn{1}{|c|}{\textbf{Output}}                 & \multicolumn{1}{l|}{3}           & \multicolumn{1}{l|}{6}           & \multicolumn{1}{l|}{3}           \\ \hline
\end{tabular}%
}
\caption{Test suite.}
\label{tab:test-suite}
\end{minipage}
\hfill
\begin{minipage}[t!]{0.44\linewidth}
\vspace{0.1in}
\centering
\begin{code}
%\caption{Fixed program returned by \Granite using the program sketch.}%
\caption{\Granite's fix using the program sketch.}%
\label{code:motivating-fixed}
\begin{minted}[escapeinside=||,tabsize=1,obeytabs,xleftmargin=2pt,linenos]{C}
int main(){
   int f,s,t;
   scanf("%d%d%d",&f,&s,&t);
   if (f >= s && f >= t)
      printf("%d",f);
   else if (s > f && s >= t)
      printf("%d",s);
   else
      printf("%d",t);

   return 0;
}
\end{minted}
\end{code}
\end{minipage}
\hfill
\end{table*}

In this paper, we propose a novel approach that combines the strengths of both FM and \LLMs to enhance APR of \IPAs via zero-shot learning. Our method involves using MaxSAT-based fault localization to identify the set of minimal buggy parts of a program and then presenting an off-the-self \LLM with a program sketch devoid of these buggy statements. This hybrid approach follows a Counterexample Guided Inductive Synthesis (\CEGIS) loop~\cite{DBLP:conf/pldi/Solar-LezamaRBE05} to iteratively refine the program. We provide the \LLM with a bug-free program sketch and ask it to synthesize the missing parts. After each iteration, the synthesized program is checked against a test suite. If the program is incorrect, a counterexample from the test suite is fed back to the \LLM, prompting a revised synthesis.

Our experiments with 1431 incorrect student programs reveal that our counterexample guided approach, utilizing MaxSAT-based bug-free program sketches, significantly boosts the repair capabilities of all six evaluated \LLMs. This method enables \LLMs to repair more programs and produce superior fixes with smaller patches, outperforming both other configurations and state-of-the-art symbolic program repair tools~\cite{clara,verifix}.
% The structure of the remainder of this paper is as follows. First, we present definitions used throughout the paper. Next, we describe our Counterexample Guided \LLM-Driven APR approach that uses hints based on MaxSAT-based Fault Localization. Then, we discuss the experimental evaluation of six state-of-the-art \LLMs for repairing \IPAs. Finally, we review related work and conclude the paper.

In summary, this paper makes the following contributions:
\begin{itemize}
\item We tackle the Automated Program Repair (APR) problem using an \LLM-Driven Counterexample Guided Inductive Synthesis (\CEGIS) approach;
\item We employ MaxSAT-based Fault Localization to guide and minimize \LLMs' patches to incorrect programs by feeding them bug-free program sketches;
\item Experiments show that our approach enables all six evaluated \LLMs to fix more programs and produce smaller patches than other configurations and symbolic tools;
\item Our code is available on GitHub~\footnote{\url{https://github.com/pmorvalho/LLM-CEGIS-Repair}} and on Zenodo~\footnote{\url{https://doi.org/10.5281/zenodo.14517771}}.
% Our code and evaluation dataset will be made open-source and publicly available on GitHub and Zenodo (see supplementary material).
\end{itemize}

% \begin{table}[t!]
%     % \vspace{-0.05in}
%     \centering
%     \setlength{\tabcolsep}{1mm}
%     \begin{tabular}{|l|*{3}{wc{1cm}|}l|wc{1.5cm}|}
%     \cline{2-4} \cline{6-6}
%     \multicolumn{1}{c|}{} & \multicolumn{3}{c|}{\textbf{Input}} &  & \textbf{Output} \\ \cline{1-4} \cline{6-6}
%     {\textbf{$t_0$}} & 1 & 2 & 3 &  & 3 \\ \cline{1-4} \cline{6-6}
%     {\textbf{$t_1$}} & 6 & 2 & 1 &  & 6 \\ \cline{1-4} \cline{6-6}
%     {\textbf{$t_2$}} & -1 & 3 & 1 &  & 3 \\ \cline{1-4} \cline{6-6}
%     \end{tabular}
% \caption{Test-suite.}
% \label{tab:test-suite}
% \end{table}

% \begin{table}[t!]
% \centering
% \label{tab:test-suite}
% % \resizebox{\columnwidth}{!}{%
% \begin{tabular}{clll}
% \cline{2-4}
% \multicolumn{1}{l|}{}                                 & \multicolumn{1}{l|}{\textit{t0}} & \multicolumn{1}{l|}{\textit{t1}} & \multicolumn{1}{l|}{\textit{t2}} \\ \hline
% \multicolumn{1}{|c|}{\multirow{3}{*}{\textbf{Input}}} & \multicolumn{1}{l|}{1}           & \multicolumn{1}{l|}{6}           & \multicolumn{1}{l|}{-1}          \\ \cline{2-4}
% \multicolumn{1}{|c|}{} & \multicolumn{1}{l|}{2} & \multicolumn{1}{l|}{2} & \multicolumn{1}{l|}{3} \\ \cline{2-4}
% \multicolumn{1}{|c|}{} & \multicolumn{1}{l|}{3} & \multicolumn{1}{l|}{1} & \multicolumn{1}{l|}{1} \\ \hline
% \multicolumn{1}{l}{}   &                        &                        &                        \\ \hline
% \multicolumn{1}{|c|}{\textbf{Output}}                 & \multicolumn{1}{l|}{3}           & \multicolumn{1}{l|}{6}           & \multicolumn{1}{l|}{3}           \\ \hline
% \end{tabular}%
% % }
% \caption{Test suite.}
% \end{table}

\section{Motivation}%
\label{sec:motivation}
Consider the program presented in Listing~\ref{code:motivating_eg}, which aims to determine the maximum among three given numbers. However, based on the test suite shown in Table~\ref{tab:test-suite}, the program is buggy as its output differs from the expected results. The set of minimal faulty lines in this program includes lines 4 and 8, as these two {\tt if} conditions are incorrect according to the test suite.
A good way to provide personalized feedback to students on their \IPAs is to highlight these two buggy lines. However, it is essential to check these faults by fixing the program and evaluating it against the test suite.

Using traditional Automated Program Repair (APR) tools for \IPAs based on Formal Methods, such as \Clara~\cite{clara} or \Verifix~\cite{verifix}, the program in Listing~\ref{code:motivating_eg} cannot be fixed within 90~seconds. \Clara takes too long to compute a `minimal' repair by considering several correct implementations for the same \IPA, while \Verifix returns a compilation error.
Conversely, using state-of-the-art \LLMs trained for coding tasks (\LLMCs), \Granite~\cite{granite-LLM-2024} or \CodeGemma~\cite{CodeGemma-LLM-2024}, would involve providing the description of the programming assignment and some examples of input-output tests. Even with these features, neither \LLM could fix the buggy program in Listing~\ref{code:motivating_eg} within 90 seconds when repeatedly testing and refining their fixes. If the lecturer's reference implementation shown in Listing~\ref{code:motivating_ref} is suggested as a reference in the prompt, both \LLMs simply copy the correct program, ignoring instructions not to do so.

Hence, symbolic approaches demand an excessive amount of time to produce an answer, and \LLMs, while fast, often produce incorrect fixes.
A promising strategy to provide feedback to students on \IPAs is to combine the strengths of both approaches. MaxSAT-based Fault localization~\cite{bugAssist-cav11,ijcai19-ignatievMWM} can rigorously identify buggy statements, which can then be highlighted in the \LLM prompt to focus on the specific parts of the program that need fixing.
Listing~\ref{code:motivating-sketch} shows an example of a program sketch, which is a partially incomplete program where each buggy statement from the original incorrect program in Listing~\ref{code:motivating_eg} is replaced with a {\tt @ HOLE @}. Instructing the \LLMs to complete this incomplete program allows both \Granite and \CodeGemma to fix the buggy program in a single interaction, returning the program in Listing~\ref{code:motivating-fixed}.

\section{Preliminaries}%
\label{sec:prelim}

This section provides definitions used throughout the paper.

\paragraph{\textbf{Synthesis Problem.}}
For a given program's specification $S$ (e.g., input-output examples), $G$ a context-free grammar (CFG), and $O$ be the semantics for a particular Domain-specific language (DSL), the goal of \emph{program synthesis} is to infer a program $\mathcal{P}$ such that (1) the program is produced by $G$, (2) the program is consistent with $O$ and (3) $\mathcal{P}$ is consistent with $S$~\cite{cp19-enum-PS,ase20-unchartIT}.
% The interest reader is referred to~\citet{compilers-book-dragon} for more context about CFGs, and DSLs.

\paragraph{\textbf{Semantic Program Repair.}}
Given ($T, G, O, P$), let $T$ be a set of input-output examples (test suite), $G$ be a grammar, $O$ be the semantics for a particular Domain-specific language, and $P$ be a syntactically well-formed program (i.e.\ sets of statements, instructions, expressions) consistent with $G$ and $O$ but semantically erroneous for at least one of the input-output tests i.e., $\exists\{t^i_{in},t^i_{out}\} \in T\ :\ P(t^i_{in}) \neq t^i_{out}$.
The goal of \emph{Semantic Program Repair} is to find a program $P_f$ by semantically change a subset $S_1$ of $P$'s statements ($S_1 \subseteq P$) for another set of statements $S_2$ consistent with $G$ and $O$, such that, $P_f = ((P \setminus S_1) \cup S_2)$ and $\forall{\{t^i_{in},t^i_{out}\} \in T}\ :\ P_f(t^i_{in}) = t^i_{out}$.

\paragraph{\textbf{Counterexample Guided Inductive Synthesis (\CEGIS).}} \CEGIS is an iterative algorithm commonly used in Program Synthesis and Formal Methods to construct programs or solutions that satisfy a given specification~\cite{cav18-CEGIS(t),jha2010oracle,DBLP:conf/pldi/Solar-LezamaRBE05}.
\CEGIS consists of two steps: the synthesis step and the verification step. Given the specification of the desired program, the inductive synthesis procedure generates a candidate program.
Next, the candidate program $P$ is passed to the verification step, which checks whether $P$ satisfies all possible inputs' specifications. Otherwise, the Decider produces a counterexample $c$ from the satisfying assignment, which is then added to the set of inputs passed to the synthesizer, and the loop repeats.
The synthesis engine refines its hypothesis using this counterexample to avoid similar mistakes in subsequent iterations. This iterative loop (comprising candidate generation, verification, counterexample generation, and refinement) continues until a correct candidate is found that satisfies all given specifications and constraints.

\paragraph{\textbf{\emph{Maximum Satisfiability (MaxSAT)}.}}
The \emph{Boolean Satisfiability} (SAT) problem is the decision problem for propositional logic~\cite{biere2009handbook}.
A propositional formula in Conjunctive Normal Form (CNF) is a conjunction of clauses where each clause is a disjunction of literals. The \emph{Maximum Satisfiability} (MaxSAT) problem is an optimization version of SAT, i.e., the goal is to find an assignment that maximizes the number of satisfied clauses in a CNF formula~\cite{sat23-UpMax}.
%the SAT problem, i.e., the goal is to find an assignment that maximizes the number of satisfied clauses in a CNF formula.
% \todo{cite MaxSAT}

\paragraph{\textbf{Formula-based Fault Localization (FBFL).}}
Given a faulty program and a test suite with failing test cases, \emph{formula-based fault localization} (FBFL) methods encode the localization problem into an optimization problem to identify a minimal set of faulty statements (diagnoses) within a program. FBFL tools
leverage MaxSAT and the theory of \emph{Model-Based Diagnosis} (MBD)~\cite{reiter87,bugAssist-pldi11,ijcai15-Marques-SilvaJI15,ijcai19-ignatievMWM,CFaults-FM24}.
% , integrating all failing test cases simultaneously~\cite{ijcai19-ignatievMWM}.
Moreover, these FBFL tools enumerate all diagnoses of a MaxSAT formula corresponding to bug locations.
% Subsequently, these methods rank diagnoses based on their frequency of appearance in each failing test.

\paragraph{Program Sketch.}
A \emph{program sketch} is a partially incomplete program where all buggy statements are replaced by placeholders, identified as ``{\tt @ HOLES @}''. These placeholders indicate parts of the program that need to be synthesized to ensure the program complies with a given specification (e.g., a test suite).
Listing~\ref{code:motivating-sketch} shows a program sketch.

\paragraph{Abstract Syntax Tree (\AST).}
An \AST is a syntax tree in which each node represents an operation, and the node's children represent the arguments of the operation for a given programming language described by a Context-Free Grammar. An \AST depicts a program's grammatical~structure~\cite{compilers-book-dragon,master-thesis-pedro}.
% \section{Approach}

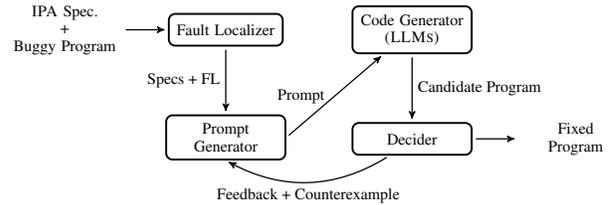
\begin{figure}[t!]
\centering
\resizebox{1\linewidth}{!}{
\begin{tikzpicture}[node distance=2cm, auto,]
 \node[punkt] (fl_node) {Fault Localizer};
 \node[punkt, inner sep=5pt,below=1.5cm of fl_node](enum) {Prompt Generator}
 edge[pil, <-] node[left=0cm of fl_node]  (l) {Specs + FL} (fl_node.south);
 \node[punkt, inner sep=5pt,right=1.5cm of fl_node]
 (llm) {Code Generator (\LLMs)}
 edge[pil, <-] node[left=0.1cm of llm]  (r) {Prompt} (enum.east);
 \node[punkt, inner sep=5pt,below=1.5cm of llm]
 (veri) {Decider} edge[pil, bend left=30] node[below] {Feedback + Counterexample} (enum.south)
 edge[pil, <-] node[right=0cm of llm]  (l) {Candidate Program} (llm.south);
\node[punkt,draw=none,left=1cm of fl_node] (specs_node) {\IPA Spec.\\+\\Buggy Program}
  edge[pil] (fl_node.west);
  \node[punkt,draw=none,right=1cm of veri] (output) {Fixed\\Program}
  edge[pil, <-] (veri.east);
\end{tikzpicture}
}
\caption{Counterexample Guided Automated Repair.}
\label{fig:cegapr}
% \vspace*{-4mm}
\end{figure}

\section{Counterexample Guided Automated Repair}

Our approach combines the strengths of both Formal Methods (FM) and \LLMs to enhance Automated Program Repair (APR). Firstly, we employ MaxSAT-based fault localization techniques to rigorously identify the minimal set of buggy parts of a program~\cite{ijcai19-ignatievMWM,CFaults-FM24}. Afterwards, we leverage \LLMs to quickly synthesize the missing parts in the program sketch. Finally, we use a counterexample from the test suite to guide \LLMs in generating patches that make the synthesized program compliant with the entire test suite, thus completing the repair.
The rationale of our approach follows a Counterexample Guided Inductive Synthesis (\CEGIS)~\cite{cegis} loop to iteratively refine the program. Figure~\ref{fig:cegapr} provides an overview of our APR approach. The input is a buggy program and the specifications for an introductory programming assignment (\IPA), including a test suite, a description of the \IPA, and the lecturer's reference implementation. We start by using MaxSAT-based fault localization techniques to identify the program's minimal set of faulty statements. Next, the prompt generator builds a prompt based on the specifications of the \IPA and a bug-free program sketch reflecting the localized faults, then feeds this information to the \LLM. The \LLM generates a program based on the provided prompt. After each iteration, the Decider module evaluates the synthesised program against a test suite. If the program is incorrect, a counterexample chosen from the test suite is sent back to the prompt generator, which then provides this counterexample to the \LLM to prompt a revised synthesis.

\tikzset{
    %Define standard arrow tip
    >=stealth',
    %Define style for boxes
    punkt/.style={
           rectangle,
           rounded corners,
           draw=black, very thick,
           text width=6.5em,
           minimum height=2em,
           text centered},
    % Define arrow style
    pil/.style={
           ->,
           thick,
           shorten <=2pt,
           shorten >=2pt,}
}

% \begin{figure}[t!]
% \centering
% \resizebox{\linewidth}{!}{
% \begin{tikzpicture}[node distance=1.5cm, auto,]
%  \node[punkt] (enum) {Prompt Generator};
%  \node[punkt, inner sep=5pt,right=0.5cm of enum]
%  (llm) {Code Generator (\LLMs)}
%  edge[pil, <-, bend right=45] node[above=0cm of llm]  (r) {Prompt} (enum.north);
%  \node[punkt, inner sep=5pt,right=0.5cm of llm]
%  (veri) {Decider} edge[pil, bend left=30] node[below] {Counterexample} (enum.south)
%  edge[pil, <-, bend right=45] node[above=0cm of llm]  (r) {Candidate Program} (llm.north);
% \node[left=0.5cm of enum] (t) {\IPA Spec.}
%   edge[pil] (enum.west);
%   \node[right=0.5cm of veri] (output) {Fixed Program}
%   edge[pil, <-] (veri.east);
% \end{tikzpicture}
% }
% \caption{Counterexample Guided Automated Repair.}
% \label{fig:cegapr}
% % \vspace*{-4mm}
% \end{figure}

\paragraph{Prompts.} The prompts fed to \LLMs can contain various types of information related to the \IPA. The typical information available in every programming course includes the description of the \IPA, the test suite to check the students' submissions corresponding to the \IPA's specifications, and the lecturer's reference implementation. The syntax used in our prompts is similar to that in other works on \LLM-driven program repair~\cite{aaai23-repair-multilingual-LLMs}.
% We have evaluated three distinct types of prompts: (1) basic prompts, (2) prompts with fault localization annotations, and (3) prompts with program sketches based on fault localization.
We have evaluated several types of prompts. Basic prompts are the simplest prompts that can be fed to an \LLM without additional computation, including all the programming assignment's basic information. An example of such a prompt is shown below:

\begin{small}
\begin{verbatim}
Fix all semantic bugs in the buggy program
below. Modify the code as little as possible.
Do not provide any explanation.

### Problem Description ###
Write a program that determines and
prints the largest of three integers
given by the user.

### Test Suite
#input:
6 2 1
#output:
6
// The other input-output tests

# Reference Implementation (Do not copy
this program) <c> #
```c
int main(){
  // Reference Implementation
}```

### Buggy Program <c> ###
```c
int main(){
  // Buggy program from Listing 1
}```

### Fixed Program <c> ###
```c
\end{verbatim}
\end{small}

In order to incorporate information about the faults localized in the program using MaxSAT-based fault localization, we utilized two different types of prompts: (1) \texttt{FIXME} annotations and (2) program sketches. \texttt{FIXME} annotated prompts are prompts where each buggy line identified by the fault localization tool is marked with a \texttt{/* FIXME */} comment. These prompts are quite similar to the basic prompt described previously, with the primary differences being the annotations in the buggy program and the first command given to the \LLMs, which is modified as follows:

\begin{small}
\begin{verbatim}
Fix all buggy lines with '/* FIXME */'
comments in the buggy program below.
\end{verbatim}
\end{small}

In the second type of prompt, to address program repair as a string completion problem, we evaluated the use of prompts where the buggy program is replaced by an incomplete program (program sketch), with each line identified as buggy by our fault localization module replaced by a \emph{hole}. The command given to the \LLMs is now to complete the incomplete program. Consequently, the sections `Buggy Program' and `Fixed Program' are replaced by `Incomplete Program' and `Complete Program', respectively, as follows:
\begin{small}
\begin{verbatim}
Complete all the '@ HOLES N @' in the
incomplete program below.
// ...
### Incomplete Program <c> ###
// ...
### Complete Program <c> ###
\end{verbatim}
\end{small}
\paragraph{Feedback.} If the candidate program generated by the \LLM is not compliant with the test suite, this feedback is provided to the \LLM in a new message through iterative querying. This new prompt indicates that the \LLM's previous suggestion to fix the buggy program was incorrect and provides a counterexample (i.e., an IO test) where the suggested fixed program produces an incorrect output. Hence, we provide the \LLM with a feedback prompt similar to:

\begin{small}
\begin{verbatim}
### Feedback ###
Your previous suggestion was incorrect!
Try again. Code only. Provide no explanation.
### Counterexample  ###
#input:
6 2 1
#output:
6

### Fixed Program <c> ###
```c
\end{verbatim}
\end{small}
% The syntax used in our prompts is similar to that in other works on \LLM-driven program repair~\cite{aaai23-repair-multilingual-LLMs,oopsla24-PyDex}.

\section{Experimental Results}
\label{sec:results}

The goal of our evaluation is to answer the following research questions: \textbf{RQ1.}  How effective are state-of-the-art (SOTA) \LLMs in repairing introductory programming assignments (\IPAs) compared to different SOTA semantic repair approaches?
%\textbf{RQ2.} Which SOTA \LLMs generate the smallest sets of program fixes?
\textbf{RQ2.} How do different prompt configurations impact the performance of \LLMs? % Ablation study
\textbf{RQ3.} How does FM-based fault localization impact \LLM-driven APR?
\textbf{RQ4.} How helpful is it to provide a reference implementation for the same \IPA to the \LLMs?
\textbf{RQ5.} What is the performance impact of using a Counterexample Guided approach in \LLM-driven APR?

\paragraph{\textbf{Experimental Setup.}}
All \LLMs were run using NVIDIA RTX A4000 graphics cards with 16GB of memory on an Intel(R) Xeon(R) Silver 4130 CPU @ 2.10GHz with 48 CPUs and 128GB RAM.
% In order to fit all \LLMs into 16GB GPUs, we used 4-bit model quantization. Moreover, all \LLMs were run using Hugging Face's Pipeline architecture.
All the experiments related to the program repair tasks were conducted on an Intel(R) Xeon(R) Silver computer with 4210R CPUs @ 2.40GHz, using a memory limit of 10GB and a timeout of 90 seconds.

\paragraph{\textbf{Evaluation Benchmark.}}
To evaluate our work, we used \benchmark~\cite{C-Pack-IPAs_apr24}, which is a set of student programs developed during an introductory programming course in the C programming language.
% To evaluate our work, we used \benchmark~\cite{C-Pack-IPAs_apr24}, which is a publicly available collection of students' programs developed during an introductory programming course in C language.
% These programs were collected over three distinct lab classes for 25 different \IPAs throughout three academic years.
% at University~X~\footnote{Temporarily removed for Double-Blind Reviewing.}.
% Each lab class focuses on a different topic of the C programming language.
% The first class deals with integers and input-output operations. Secondly, the second class focuses on loops and chars. Lastly, in the third lab class, students learn to use vectors and strings.
% The textual description of each programming assignment can be found in the supplementary material.
% Appendix~\ref{appendix:description-IPAs}.
% This dataset of introductory programming exercises, \benchmark\cite{C-Pack-IPAs_apr24}, is publicly available.
% Moreover, \benchmark~\cite{C-Pack-IPAs,C-Pack-IPAs_apr24} has also proven successful in evaluating various works across program analysis~\cite{aitp22-learning-2-map-variables,ecai23-GNNs-4-var-mapping}, and program transformation~\cite{fse22-MultIPAS}.
Since this work focuses only on semantic program repair, only programs that compile without any errors were selected.
% The set of submissions was split into two sets: correct submissions and incorrect submissions.
% The students' submissions that satisfied a set of input-output test cases for each \IPA were considered correct.
\benchmark contains 1431 semantically incorrect programs, i.e., fail at least one input-output test.
% submitted for 25 different \IPAs.
% This dataset of \IPAs will be publicly available.

\paragraph{Large Language Models (\LLMs).}
% \label{sec:llms}
% In our experiments, we chose to use only open-access models available on Hugging Face~\cite{huggingface} with approximately 7 billion parameters for two primary reasons. Firstly, closed-access models like Chat-GPT are cost-prohibitive and raise concerns over student data privacy. Secondly, models with a very large number of parameters necessitate significant computational resources, such as GPUs with higher RAM capacities, and take longer to generate responses, which is not suitable for a classroom setting.
In our experiments, we used only open-access \LLMs available on Hugging Face~\cite{huggingface} with approximately 7 billion parameters for three primary reasons. Firstly, closed-access models like Chat-GPT are cost-prohibitive and raise concerns over student data privacy. Secondly, models with a very large number of parameters (e.g., 70B) need significant computational resources, such as GPUs with higher RAM capacities, and take longer to generate responses, which is unsuitable for a classroom setting. Thirdly, we used these off-the-shelf \LLMs to evaluate the publicly available versions without fine-tuning them. This approach ensures that the \LLMs used in this paper are available to anyone without investing time and resources into fine-tuning these models.
% With these considerations in mind
Thus, we evaluated six different \LLMs for this study through iterative querying. Three of these models are \LLMCs, i.e., \LLMs fine-tuned for coding tasks: IBM's \Granite~\cite{granite-LLM-2024}, Google's \CodeGemma~\cite{CodeGemma-LLM-2024} and Meta's \CodeLlama~\cite{CodeLlama-LLM-2023}. The other three models are general-purpose \LLMs not specifically tailored for coding tasks: Google's \Gemma~\cite{Gemma-LLM-2024}, Meta's \Llama (latest version of the \textsc{Llama} family~\cite{Llama-LLMs-2023}) and Microsoft's \lPhi~\cite{Phi3-LLM-2024}.
% This selection allowed us to explore a balance between accessibility, performance, and relevance to the domain of program repair within an educational context.

% Regarding each model specification, we used \Llama's 8B-parameter instruction tuned variant.
% \CodeLlama 7B-instruct-tuned version designed for general code synthesis and understanding.
% Furthermore, we used \Granite's 8B-parameter model fine tuned to respond to coding related instructions.
% We used \lPhi's 3.8 billion-parameter mini version with 128K context length.
% Finally, we used the 7B-instruct version of the \Gemma model, and \CodeGemma's 7B-parameter instruction-tuned variant for code chat and instruction.

We selected specific variants of each model to optimize their performance for our program repair tasks. For Meta's \Llama, we utilized the 8B-parameter instruction-tuned variant. This model is designed to follow instructions more accurately, making it suitable for a range of tasks, including program repair. For \CodeLlama, we used the 7B-parameter instruct-tuned version, which is specifically designed for general code synthesis and understanding, making it highly effective for coding tasks.
We employed \Granite model with 8B-parameters, fine-tuned to respond to coding-related instructions. For \lPhi, we opted for the mini version, which has 3.8B-parameters and a context length of 128K. This smaller model is efficient yet capable of handling extensive context, making it practical for educational settings.
For \Gemma, we used the 7B-parameter instruction-tuned version, optimized to follow detailed instructions. Lastly, for \CodeGemma, we selected the 7B-parameter instruction-tuned variant, designed specifically for code chat and instruction, enhancing its capability to handle programming-related queries and tasks. To fit all \LLMs into 16GB GPUs, we used model quantization of 4bit. Moreover, all \LLMs were run using Hugging Face's Pipeline architecture.
By using these different \LLMs, we aimed to balance computational efficiency with the ability to effectively generate and refine code, facilitating a practical APR approach in an educational environment.

% \subsection{Fault Localization}
% \label{sec:fault_loc}

\paragraph{Fault Localization (FL).} We used \textsc{CFaults}~\cite{CFaults-FM24}
% F~\footnote{Temporarily removed for Double-Blind Reviewing.},
which is a formula-based FL tool, that pinpoints bug locations within the programs. It aggregates all failing test cases into a unified MaxSAT formula. This FL tool can be easily replaced by other FL tools.
% (e.g. spectrum-based).
% (see Section~ \ref{sec:prelim} ).

\begin{table*}[t!]
\centering
% \begin{tabular}{|c|cccccc|}
\resizebox{2\columnwidth}{!}{
\begin{tabular}{@{\extracolsep{8pt}}ccccccccc}
\toprule
% {} & \multicolumn{7}{c}{\textbf{Entire Benchmark}} \\ \hline
{} & \multicolumn{7}{c}{\textbf{Configurations without access to Reference Implementations}}  \\ \hline
\textbf{\LLMs}  & \textbf{De-TS}  & \textbf{De-TS-CE}  & \textbf{FIXME\_De-TS}  & \textbf{FIXME\_De-TS-CE}  & \textbf{Sk\_De-TS}  & \textbf{Sk\_De-TS-CE}  & \textbf{\begin{tabular}[c]{@{}c@{}}Portfolio\\(All Configurations)\end{tabular}} \\
\hline
\textbf{CodeGemma}  & 597 (41.7\%)  & 606 (42.3\%)  & 592 (41.4\%)  & 601 (42.0\%)  & 682 (47.7\%)  & \textbf{688 (48.1\%)} & 823 (57.5\%)\\
\textbf{CodeLlama}  & 492 (34.4\%)  & 500 (34.9\%)  & 481 (33.6\%)  & 463 (32.4\%)  & \textbf{573 (40.0\%)}  & 561 (39.2\%) & 712 (49.8\%)\\
\textbf{Gemma}  & 496 (34.7\%)  & 492 (34.4\%)  & 446 (31.2\%)  & 444 (31.0\%)  & 532 (37.2\%)  & \textbf{534 (37.3\%)} & 670 (46.8\%)\\
\textbf{Granite}  & 626 (43.7\%)  & 624 (43.6\%)  & 566 (39.6\%)  & 583 (40.7\%)  & \textbf{691 (48.3\%)}  & 681 (47.6\%) & 846 (59.1\%)\\
\textbf{Llama3}  & 564 (39.4\%)  & 590 (41.2\%)  & 535 (37.4\%)  & 557 (38.9\%)  & 578 (40.4\%)  & \textbf{591 (41.3\%)} & 851 (59.5\%)\\
\textbf{Phi3}  & 494 (34.5\%)  & 489 (34.2\%)  & 460 (32.1\%)  & 474 (33.1\%)  & \textbf{547 (38.2\%)}  & 535 (37.4\%) & 621 (43.4\%)\\
\midrule
\textbf{\begin{tabular}[c]{@{}c@{}}Portfolio\\(All \LLMs)\end{tabular}} & 842 (58.8\%)  & 846 (59.1\%)  & 796 (55.6\%)  & 820 (57.3\%)  & 900 (62.9\%)  & \textbf{907 (63.4\%)}  & 1013 (70.8\%)& \\
% \midrule
% \textbf{Verifix}  & 90 (6.3\%) &  &  &  &  &  & \\
% \textbf{Clara}  & 495 (34.6\%) &  &  &  &  &  & \\
\bottomrule
 \end{tabular}} \\
\vspace{0.1in}
\resizebox{2\columnwidth}{!}{
\begin{tabular}{@{\extracolsep{8pt}}ccccccccc}
\toprule
% {} & \multicolumn{7}{c}{\textbf{Entire Benchmark}} \\ \hline
{} & \multicolumn{7}{c}{\textbf{Configurations with access to Reference Implementations}}  \\ \hline
\textbf{\LLMs}  & \textbf{De-TS-CE-CPA}  & \textbf{De-TS-CE-RI}  & \textbf{FIXME\_De-TS-CE-CPA}  & \textbf{FIXME\_De-TS-CE-RI}  & \textbf{Sk\_De-TS-CE-CPA}  & \textbf{Sk\_De-TS-CE-RI}  & \textbf{\begin{tabular}[c]{@{}c@{}}Portfolio\\(All Configurations)\end{tabular}} \\
\hline
\textbf{CodeGemma}  & 578 (40.4\%)  & 576 (40.3\%)  & 637 (44.5\%)  & 638 (44.6\%)  & 725 (50.7\%)  & \textbf{739 (51.6\%)} & 916 (64.0\%)\\
\textbf{CodeLlama}  & 528 (36.9\%)  & 525 (36.7\%)  & 565 (39.5\%)  & 609 (42.6\%)  & 633 (44.2\%)  & \textbf{675 (47.2\%)} & 893 (62.4\%)\\
\textbf{Gemma}  & 595 (41.6\%)  & 607 (42.4\%)  & 563 (39.3\%)  & 616 (43.0\%)  & 664 (46.4\%)  & \textbf{732 (51.2\%)} & 951 (66.5\%)\\
\textbf{Granite}  & 773 (54.0\%)  & 828 (57.9\%)  & 794 (55.5\%)  & 857 (59.9\%)  & 838 (58.6\%)  & \textbf{876 (61.2\%)} & 1132 (79.1\%)\\
\textbf{Llama3}  & 685 (47.9\%)  & 691 (48.3\%)  & 657 (45.9\%)  & 681 (47.6\%)  & 725 (50.7\%)  & \textbf{730 (51.0\%)} & 1016 (71.0\%)\\
\textbf{Phi3}  & 552 (38.6\%)  & 444 (31.0\%)  & 545 (38.1\%)  & 492 (34.4\%)  & 639 (44.7\%)  & \textbf{647 (45.2\%)} & 899 (62.8\%)\\
\midrule
\textbf{\begin{tabular}[c]{@{}c@{}}Portfolio\\(All \LLMs)\end{tabular}} & 1033 (72.2\%)  & 1046 (73.1\%)  & 1011 (70.6\%)  & 1056 (73.8\%)  & 1050 (73.4\%)  & \textbf{1077 (75.3\%)}  & 1190 (83.2\%)& \\
% \midrule
% \textbf{Verifix}  & 90 (6.29\%) &  &  &  &  &  & \\
% \textbf{Clara}  & 495 (34.59\%) &  &  &  &  &  & \\
\bottomrule
 \end{tabular}} \\
% \vspace{0.2in}
\caption{
The number of programs fixed by each \LLM under various configurations. Row \textit{Portfolio (All \LLMs)}, shows the best results across all \LLMs for each configuration. Column \textit{Portfolio (All Configurations)} shows the best results for each \LLM across all configurations.
% The last two rows show the performance of \Clara and \Verifix on our evaluation benchmark.
Mapping abbreviations to configuration names:
\textbf{De} - \textit{\IPA Description},
\textbf{TS} - \textit{Test Suite},
\textbf{CE} - \textit{Counter\-example},
\textbf{RI} - \textit{Reference Implementation},
\textbf{CPA} - \textit{Closest Program using \ASTs},
\textbf{FIXME} - \textit{FIXME Annotations},
\textbf{SK}~-~\textit{Sketches}.
% \textbf{All} includes \textbf{ID}, \textbf{IO}, \textbf{CE}, and \textbf{RI}. If "No Ref Implementation" is present, \textbf{All} excludes \textbf{RI}.
% \textbf{SK} and \textbf{FIXME} are prioritized at the beginning.
}
\label{tab:repair}
\end{table*}

\begin{table*}[t!]
\centering
% \begin{tabular}{|c|cccccc|}
\resizebox{2\columnwidth}{!}{
\begin{tabular}{@{\extracolsep{8pt}}ccccccccccc}
\toprule
% {} & \multicolumn{8}{c}{\textbf{Entire Benchmark}} \\ \hline
{} & \multicolumn{8}{c}{Metric: \textbf{sum(Distance Score)}}  \\ \hline
{} & \multicolumn{8}{c}{\textbf{Configurations}}  \\ \hline
\textbf{\LLMs}  & \textbf{De-TS}  & \textbf{De-TS-CE}  & \textbf{De-TS-CE-CPA}  & \textbf{De-TS-CE-RI}  & \textbf{Sk\_De-TS}  & \textbf{Sk\_De-TS-CE}  & \textbf{Sk\_De-TS-CE-CPA}  & \textbf{Sk\_De-TS-CE-RI} \\
\hline
\textbf{CodeGemma}  & 471.0  & 486.4  & 429.7  & 440.4  & 524.4  & \textbf{529.5}  & 249.8  & 497.3 & \\
\textbf{CodeLlama}  & 437.5  & 438.8  & 409.5  & 404.8  & \textbf{477.9}  & 464.5  & 251.3  & 459.0 & \\
\textbf{Gemma}  & 306.5  & 296.9  & \textbf{370.8}  & 231.0  & 338.8  & 340.3  & 156.4  & 316.2 & \\
\textbf{Granite}  & 512.8  & 506.3  & 453.4  & 292.1  & \textbf{539.8}  & 533.6  & 172.3  & 334.5 & \\
\textbf{Llama3}  & 367.9  & 368.0  & 414.8  & 381.9  & 379.8  & 384.5  & 172.7  & \textbf{423.0} & \\
\textbf{Phi3}  & 291.9  & 292.6  & 287.6  & 148.1  & \textbf{326.5}  & 321.4  & 98.2  & 253.4 & \\
\bottomrule
 \end{tabular}
 }
 \\
%  \resizebox{2\columnwidth}{!}{
% \begin{tabular}{@{\extracolsep{8pt}}cccccccc}
% \toprule
% {} & \multicolumn{5}{c}{\textbf{Entire Benchmark}} \\ \hline
% {} & \multicolumn{5}{c}{Metric: \textbf{sum(Distance Score)}}  \\ \hline
% {} & \multicolumn{5}{c}{\textbf{Configurations}}  \\ \hline
% \textbf{\LLMs}  & \textbf{De-TS-CE-CPA}  & \textbf{De-TS-CE-RI}  & \textbf{FIXME\_De-TS-CE-CPA}  & \textbf{RI}  & \textbf{Sk\_De-TS-CE-CPA} \\
% \hline
% \textbf{CodeGemma}  & 429.7  & \textbf{440.37}  & 249.21  & 373.53  & 249.82 & \\
% \textbf{CodeLlama}  & \textbf{409.54}  & 404.76  & 240.44  & 332.34  & 251.27 & \\
% \textbf{Gemma}  & \textbf{370.85}  & 230.99  & 142.21  & 227.25  & 156.39 & \\
% \textbf{Granite}  & \textbf{453.35}  & 292.13  & 171.15  & 93.78  & 172.31 & \\
% \textbf{Llama3}  & \textbf{414.75}  & 381.9  & 175.91  & 409.81  & 172.67 & \\
% \textbf{Phi3}  & \textbf{287.57}  & 148.06  & 96.81  & 203.22  & 98.16 & \\
% \bottomrule
%  \end{tabular}} \\
\caption{
The cumulative distance scores for each program successfully repaired by each \LLM across various configurations.
}
\label{tab:distance-score}

\end{table*}

\begin{table*}[t!]
\centering
% \begin{tabular}{|c|cccccc|}
\resizebox{2\columnwidth}{!}{
\begin{tabular}{@{\extracolsep{8pt}}ccccccc}
\toprule
\textbf{\LLMs}  &  \textbf{CodeGemma+Sk\_De-TS-CE} &  \textbf{CodeLlama+Sk\_De-TS} & \textbf{Gemma+Sk\_De-TS-CE} & \textbf{Granite+Sk\_De-TS} & \textbf{Llama3+Sk\_De-TS-CE} & \textbf{Phi3+Sk\_De-TS} \\
\hline
\textbf{\#Programs Fixed} &  270 (34.5\%) & 210 (26.8\%) & 210 (26.8\%) & \textbf{290 (37.0\%)} & 199 (25.4\%) & 199 (25.4\%) \\
\bottomrule
 \end{tabular}
 }
\caption{
The number of programs repaired by each \LLM using their best-performing prompt configuration, specifically on the subset of programs where \Clara fails to repair due to control-flow issues (54.7\% of \benchmark).
}
\label{tab:progs-clara-fails}

\end{table*}

\begin{table*}[t!]
\centering
% \begin{tabular}{|c|cccccc|}
\resizebox{2\columnwidth}{!}{
\begin{tabular}{@{\extracolsep{8pt}}cccccccc}
\toprule
% \textbf{Quartiles}
\textbf{\begin{tabular}[c]{@{}c@{}}Quartiles for Average\\Cyclomatic Complexity\end{tabular}}
&  \textbf{CodeGemma+Sk\_De-TS-CE} &  \textbf{CodeLlama+Sk\_De-TS} & \textbf{Gemma+Sk\_De-TS-CE} & \textbf{Granite+Sk\_De-TS} & \textbf{Llama3+Sk\_De-TS-CE} & \textbf{Phi3+Sk\_De-TS}  &  \textbf{\Clara} \\
\hline
\textbf{Q1: 1.0-2.5} &  \textbf{231 (76.5\%)} & 178 (58.9\%) & 183 (60.6\%) & 201 (66.6\%) & 217 (71.9\%) & 198 (65.6\%)  & 164 (54.3\%) \\
\textbf{Q2: 2.5-3.5} &  \textbf{231 (65.3\%)} & 212 (59.9\%) & 169 (47.7\%) & \textbf{231 (65.3\%)} & 192 (54.2\%) & 180 (50.9\%) & 163 (46.1\%) \\
\textbf{Q3: 3.5-7.0} &  168 (45.3\%) & 140 (37.7\%) & 129 (34.8\%) & \textbf{190 (51.2\%)} & 139 (37.5\%) & 125 (33.7\%) & 124 (33.4\%) \\
\textbf{Q4: 7.0-26} &  58 (14.4\%) & 43 (10.6\%) & 53 (13.1\%) & \textbf{69 (17.1\%)} & 43 (10.6\%) & 44 (10.9\%) & 44 (10.9\%) \\

\bottomrule
 \end{tabular}
 }
\caption{
The number of programs repaired by each \LLM using their best-performing prompt configuration, considering the average cyclomatic complexity of programs~\cite{lizard}.}
\label{tab:cyclomatic-complexity}

\end{table*}

\subsection{Evaluation.}
To assess the effectiveness of the program fixes generated by the \LLMs under different prompt configurations, we used two key metrics: the number of programs successfully repaired and the quality of the repairs.
% the time performance of each model,
% For assessing the quality of the repairs, we computed \emph{tree edit distance (TED)} that measures the distance between two programs' \ASTs.
% two metrics that measure the distance between two programs' \ASTs: \emph{tree edit distance (TED)} and \emph{Number of Matching Sub-ASTs (NMS)}.
For assessing the patch quality, we use the \emph{Tree Edit Distance} (TED)~\cite{tree-edit-distance79,tree-edit-distance-jc89} to compute the distance between the student's buggy program and the fixed program returned by the \LLMs.
TED computes the structural differences between two Abstract Syntax Trees (\ASTs) by calculating the minimum number of edit operations (i.e., insertions, deletions, and substitutions) needed to transform one \AST into another.
% \paragraph{Number of Matching Sub-ASTs (NMS).}
% We also count the \emph{Number of Matching Sub-ASTs (NMS)}.
% Computing subtree matching involves identifying and counting the number of matching subtrees between two \ASTs. It detects common code structures, and by quantifying matching subtrees, we gain insights into the similarity between two programs.
% Based on these two metrics for measuring program distances, TED and MSA, we computed the following scores: \emph{distance score} and \emph{matching score}.
Based on this metric for measuring program distances, we computed the \emph{distance score}, defined by Equation~\ref{eq:distance-score}. This score aims to identify and penalize \LLMs that replace the buggy program with the reference implementation rather than fixing it. The distance score is zero when the TED of the original buggy program ($T_{o}$) to the program suggested by the \LLM ($T_{f}$) is the same as the TED of the reference implementation ($T_{r}$) to $T_{o}$. Otherwise, it penalizes larger fixes than necessary to align the program with the correct implementation.

\begin{equation}
    \label{eq:distance-score}
    ds(T_{f},T_{o},T_{r})=max\Big(0, 1-\frac{\ted(T_{f},T_{o})}{\ted(T_{r},T_{o})}\Big)
\end{equation}

% The \emph{matching score} is defined by Equation~\ref{eq:matching-score} and measures the similarity between the fixed program ($T_{f}$) and the original buggy program ($T_{o}$) based on the number of matching sub-\ASTs. This score is calculated as the ratio of the number of matching sub-ASTs between $T_{f}$ and $T_{o}$ to the number of matching sub-\ASTs within $T_{o}$ itself. This metric helps assess how much of the original program's structure is retained in the fixed version returned by the \LLMs, with higher scores indicating more structural similarity.

% \begin{equation}
%     \label{eq:matching-score}
%     ms(T_{f},T_{o})=\frac{\#MatchingSubASTs(T_{f},T_{o})}{\#MatchingSubASTs(T_{o},T_{o})}
% \end{equation}

% \subsubsection{Dead code.}

\paragraph{Baseline.} We used two state-of-the-art traditional semantic program repair tools for \IPAs as baselines: \Verifix~\cite{verifix} and \Clara~\cite{clara}. \Verifix employs MaxSMT to align a buggy program with a reference implementation provided by the lecturer, while \Clara clusters multiple correct implementations and selects the one that produces the smallest fix when aligned with the buggy program. Both tools require an exact match between the control flow graphs (e.g., branches, loops) and a bijective relationship between the variables; otherwise, they return a structural mismatch error. \Verifix was provided with each buggy program, the reference implementation, and a test suite. \Clara was given all correct programs from different academic years to generate clusters for each \IPA.
% With a time limit of 90 seconds, \Verifix can only repair 6.3\% of the benchmark due to structural and unsupported errors, while \Clara repairs 34.6\%.
Within a 90-second time limit, \Clara repairs 495 programs (34.6\%), times out without producing a repair on 154 programs (10.8\%), and fails to repair 738 programs (54.7\%). In comparison, \Verifix repairs 91 programs (6.3\%), reaches the time limit on 0.6\%, and fails to repair 1338 programs (93.5\%). The main reason for these failures is that both tools rely on structure mismatch errors.

Table~\ref{tab:repair} presents the number of programs successfully repaired by each \LLM under various configurations. The row labeled \texttt{Portfolio} represents the best possible outcomes by selecting the optimal configuration for each program across all \LLMs. Meanwhile, \texttt{Portfolio} column highlights the best results achieved by a particular \LLM across all tested configurations.
The configurations yielding the highest success rates for the six evaluated \LLMs involve incorporating a reference implementation of the \IPA into the prompt. However, rather than genuinely fixing the buggy program, the \LLMs often replace it with the reference implementation. For instance, \Granite repairs 876 programs using a configuration that includes bug-free program sketches (Sk), an \IPA description, counterexamples, a test suite, and the reference implementation (Sk\_De-TS-CE-RI). Notably, 442 of these repaired programs exhibit a TED value of zero between the reference implementation and the fixed program, indicating that \Granite is replicating the reference implementation.
To address this, we separately analyzed configurations that include and exclude access to a reference implementation.
When no reference implementation is provided (top of Table~\ref{tab:repair}), \Granite still leads among the \LLMs, fixing up to 59.1\% of the programs across all configurations and 48.3\% when using sketches (SK), the \IPA description, and a test suite (SK\_De-TS). \CodeGemma also performs well, achieving up to 57.5\% success in a portfolio approach and showing particular strength in configurations involving sketches (SK). For instance, \CodeGemma can repair 48.1\% of the evaluation benchmark using bug-free sketches, \IPA description, test suite, and counterexample (SK\_De-TS-CE).
Configurations incorporating sketches (SK) and FIXME annotations generally yield better results. Including counterexamples (CE), \IPA descriptions, and test suites (De-TS) further boosts the success rate across different \LLMs.
% Traditional APR tools like \Clara and \Verifix, however, are less effective, with Clara fixing 34.6\% of programs and Verifix only 6.3\%. On the other hand,
The portfolio approach, which combines the strengths of all \LLMs and configurations without using reference implementation, achieves the highest overall success rate, fixing 70.8\% of the programs. This demonstrates that leveraging multiple \LLMs together can significantly enhance repair success.

Moreover, considering only CEGIS loops where \LLMs were able to repair the program within the time limit, the minimum number of iterations is one, the maximum number of iterations to fix a program is seven, and the average number of iterations is 1.14. In 89\% of the cases, the program is repaired on the first attempt.

\begin{figure}[t!]
    \centering
    \includegraphics[width=0.635\linewidth]{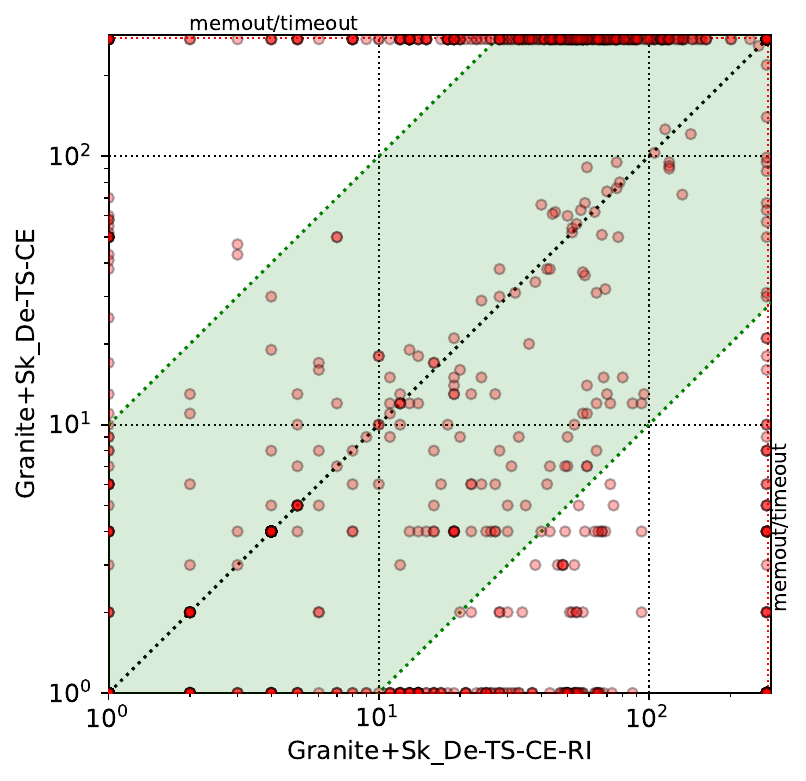}
    \caption{Comparison of tree edit distances~(TED) for \Granite's repairs when using (x-axis) versus not using (y-axis) correct implementations with configuration~Sk\_De-TS-CE.}
    \label{fig:scatter-plot}
\end{figure}

Furthermore, we provide the results of \LLMs with a reference implementation (bottom of Table~\ref{tab:repair}). The reference implementation can be either the lecturer's implementation for the same \IPA or the closest correct program based on the programs' Abstract Syntax Trees (ASTs) from a previously submitted student program, determined by Tree Edit Distance (TED) values~\cite{InvAASTCluster-corr22}. The intent was to allow the model to reuse correct code snippets to generate repairs. Results show that including a reference implementation allows for better repair results. However, as mentioned earlier, the \LLMs often simply copy the provided reference implementation.

Table~\ref{tab:distance-score} presents the sum of the distance scores (see Eq.~\ref{eq:distance-score}) for the top-performing \LLMs from Table~\ref{tab:repair} across different configurations. This summation aims to penalize \LLMs that either copy the provided reference implementation or generate unnecessarily large repairs. For example, \Granite using configuration Sk\_De-TS-CE-RI can repair 876 programs but yields a total distance score of 334.5, whereas using the same configuration without a correct implementation repairs 681 programs resulting in a higher distance score of 533.6.
%Figure~\ref{fig:scatter-plot} presents a scatter plot that compares the TED values achieved by \Granite when provided with a reference implementation versus when it operates without one using configuration Sk\_De-TS-CE. Each point on this plot represents a faulty program, where the x-value (resp. y-value) represents the TED value of \Granite' with access to a reference implementation (resp. without access to it). Points below the diagonal indicate that fixing a program with access to a correct implementation incurs a higher TED cost than fixing it without access. This suggests that while access to a reference implementation enables \Granite and other \LLMs to repair more programs, it often results in larger modifications to the student's program than when no reference implementation is given.

Figure~\ref{fig:scatter-plot} shows a scatter plot that compares the tree edit distance (TED) of the buggy program to the program fixed by \Granite with and without a reference implementation, using configuration Sk\_De-TS-CE. Each point represents a faulty program, where the x-value (resp. y-value) represents the TED cost of \Granite' with access to a reference implementation (resp. without it). Points below the diagonal indicate that fixing a program with access to a correct implementation incurs a higher TED cost than fixing it without access. This suggests that while access to a reference implementation enables \Granite and other \LLMs to repair more programs, it often results in larger changes to the student's program than when no correct implementation is given.

%To answer our research questions: For RQ1, all six \LLMs using FL-based sketches, \IPA descriptions, test suite, and counterexamples repair more programs than traditional repair tools. Regarding RQ2, although \LLMs generate numerous fixes, including a reference implementation, as explored in RQ4, often results in larger and potentially less efficient fixes. For RQ3, prompt configurations incorporating FL-based Sketches, FIXME annotations, and Counterexamples yield the most successful repair outcomes. Finally, in response to RQ6, employing a Counterexample guided approach significantly improves the accuracy of \LLM-driven program repair across various configurations.

\subsection{Discussion.}

To answer our research questions: For RQ1, all six \LLMs using different prompt configurations repair more programs than traditional repair tools. For RQ2, prompt configurations with FL-based Sketches, \IPA description and test suite yield the most successful repair outcomes. Moreover, for RQ3, it is clear that incorporating FL-based Sketches (or even FIXME annotations) allows the \LLMs to repair more programs than only providing the buggy program. For RQ4, including a reference implementation allows for more repaired programs but with potentially less efficient fixes. Finally, for RQ5, employing a Counterexample guided approach significantly improves the accuracy of \LLM-driven APR across various configurations.
Counterexamples help in the repair process of certain \LLMs, such as \CodeGemma and \Llama, across all prompt configurations. For other \LLMs, counterexamples are beneficial but only in specific configurations. This difference may be due to variations in the training data used for each \LLM.
However, a more detailed analysis is necessary.

Furthermore, we analyzed the effectiveness of \LLMs in repairing programs that \Clara fails to address due to control-flow issues, representing 54.7\% of \benchmark~(738 programs). Table~\ref{tab:progs-clara-fails} presents these results. Among the best-performing configurations, \Granite with Sk\_De-TS achieved the highest repair rate, successfully fixing 290 programs (37.0\%) in this subset. This highlights \Granite's strong capability to handle complex program structures where traditional constraint-based tools fail. \CodeGemma with Sk\_De-TS-CE also performed well, repairing 270 programs (34.5\%), demonstrating the advantage of incorporating counterexamples (CE) alongside the Sketches (Sk) configuration. In contrast, models such as \Llama and \lPhi achieved lower success rates, each repairing only 199 programs (25.4\%), suggesting limitations in their ability to generalize and address intricate control-flow issues.

To gain deeper insights into \LLMs' performance across varying levels of program complexity, we evaluated the average cyclomatic complexity of each program in \benchmark using \texttt{lizard}~\cite{lizard}. Table~\ref{tab:cyclomatic-complexity} summarizes these findings, divided into quartiles based on cyclomatic complexity. For simpler programs (Q1: 1.0–2.5), \CodeGemma+Sk\_De-TS-CE excelled, achieving a 76.5\% repair rate. However, as program complexity increased (Q3: 3.5–7.0), \Granite+Sk\_De-TS outperformed the other models with a 51.2\% repair rate, underscoring its robustness in tackling moderately complex programs. In the most challenging cases (Q4: 7.0–26), \Granite retained its lead, repairing 17.1\% of programs. These results suggest that while \CodeGemma is highly effective for simpler errors, \Granite exhibits superior adaptability and resilience when addressing programs of greater complexity.
Table~\ref{tab:cyclomatic-complexity} also shows that all evaluated models, including \Clara, face significant challenges in repairing programs with an average cyclomatic complexity higher than seven.

\section{Related Work}

Several constraint-based program repair techniques have been proposed to check if a student's program is semantically correct:
% semantic-based~\cite{semFix, directFix, angelix},
clustering-based~\cite{clara},
implementation-driven~\cite{sarfgen, verifix, refactory, autograder}, and semantic code search~\cite{sosRepair-tse19}.
% static analysis violations~\cite{phoenix-fse, phoenix-icse},
% and generate-and-validate techniques~\cite{genProg, prophet-popl16}.
% \emph{Solution-driven program repair} can be divided into two main approaches clustering-based and implementation-based.
\emph{Clustering-based repair} tools~\cite{clara} receive an incorrect program, a test suite, and a set of correct student submissions for the same \IPA.
\emph{Implementation-driven repair} tools use one reference implementation to repair a given incorrect submission~\cite{verifix}.

Large Language Models (\LLMs) trained on code (\LLMCs) have demonstrated significant effectiveness in generating program fixes~\cite{aaai23-repair-multilingual-LLMs,ase23-LLMs-plastic-surgery,fse23-copiloting-copilots,icse23-APR-LLMs,icse23-APR-pretrained-LLMs,icst25_LLM-based_Repair_ASPs,tosem24-BatFix}.
For instance, \textsc{RING}~\cite{aaai23-repair-multilingual-LLMs} is a multilingual repair engine powered by an \LLMC that uses fault localization (FL) information from error messages and leverages the few-shot capabilities of \LLMCs for code transformation.
In the context of Automated Program Repair (APR) for programming education, several works have explored the use of \LLMs for coding tasks~\cite{oopsla24-PyDex,edm23-feedback-syntax-errors-LLMs,codeHelp-koli23}. PyDex~\cite{oopsla24-PyDex}, for example, employs iterative querying with \textsc{Codex}, an \LLMC version of ChatGPT, using test-based few-shot selection and structure-based program chunking to repair syntax and semantic errors in Python assignments. Similarly, \textsc{CodeHelp}~\cite{codeHelp-koli23} utilizes OpenAI's \LLMs to provide textual feedback to students on their assignments. However, to the best of our knowledge, no existing work has explored the use of \LLMs guided by formula-based FL.

\section{Conclusion}

Large Language Models (\LLMs) excel at completing strings, while MaxSAT-based fault localization (FL) excels at identifying buggy parts of a program. We proposed a novel approach combining MaxSAT-based FL and \LLMs via zero-shot learning to enhance Automated Program Repair (APR) for introductory programming assignments (\IPAs).
% Our method identifies buggy statements, presents the \LLM with a program sketch devoid of these faults, and asks it to synthesize the missing parts, which are then checked against a test suite. Incorrect suggestions are revised using counterexamples from the test suite.
Experiments show that our bug-free program sketches significantly improve the repair capabilities of all six evaluated \LLMs, enabling them to repair more programs and produce smaller patches compared to other configurations and state-of-the-art symbolic program repair tools. Therefore, this interaction between Formal Methods and \LLMs yields more accurate and efficient program fixes, enhancing feedback mechanisms in programming education.

\section{Acknowledgments}

PO acknowledges support from the EU’s Horizon 2020 research and innovation programme under ELISE Grant Agreement No 951847 and the ERC AdG FUN2MODEL (Grant agreement No. 834115).
This work was partially supported by Portuguese national funds through FCT, under projects UIDB/50021/2020 (DOI: 10.54499/\-UIDB/\-50021/\-2020), PTDC/\-CCI-COM/\-2156/2021 (DOI: 10.54499/\-PTDC/\-CCI-COM/\-2156/\-2021) and 2023.14280.PEX (DOI: 10.54499/2023.14280.PEX) and grant SFRH/\-BD/\-07724/\-2020 (DOI: 10.54499/\-2020.07724.BD).
This work was also supported by the MEYS within the program ERC CZ under the project POSTMAN no.~LL1902 and co-funded by the EU under the project \emph{ROBOPROX} (reg.~no.~CZ\-.02.01.01/00/\-22\_008/0004590). 
\bibliography{mybibliography}

\clearpage

% \appendix
% \input{chklst}

% \clearpage

% \input{appendix}

\end{document}